\documentstyle[epsfig]{aipproc}
\newcommand{\be}{\begin{equation}}
\newcommand{\ee}{\end{equation}}
\newcommand{\bea}{\begin{eqnarray}}
\newcommand{\eea}{\end{eqnarray}}

\newcommand{\p}{\partial}
\newcommand{\R}{{\rm l \! R}}

\begin{document}
\title{Phase Transition in Quantum Gravity}

\author{Viqar Husain and Sebastian Jaimungal}
\address{Department of Physics and Astronomy,\\
University of British Columbia,\\ 
6224 Agricultural Road, Vancouver, BC V6T~1Z1, Canada}

%\lefthead{LEFT head}
%\rig436thead{RIGHT head}
\maketitle

\begin{abstract}
A fundamental problem with attempting to quantize general relativity
is its perturbative non-renormalizability. However, this fact does not
rule out the possibility that non-perturbative effects can be
computed, at least in some approximation. We outline a quantum field
theory calculation, based on general relativity as the classical
theory, which implies a phase transition in quantum gravity. The 
order parameters are composite fields derived from spacetime metric
functions. These are massless below a critical energy scale and become
massive above it. There is a corresponding breaking of classical symmetry.
\end{abstract}

A quantum theory of gravity should have the property that its long
distance regime is classical general relativity (GR).  On the other
hand, the only criteria for the short distance regime, in the absence
of experimental guides, is self-consistency and finiteness.  At first
sight this appears to bestow a large degree of freedom on attempts at
formulating a theory of quantum gravity. However, as is well known,
there have been many unsuccessful attempts over the years, with partial
successes coming mainly from string theory \cite{strings}, and the 
loop quantum gravity program \cite{loop}. 

From a conventional standpoint, a first question might be what
classical theory is to be quantized to obtain quantum gravity. An
obvious starting point, classical GR, leads to the
non-renormalizability problem as far as perturbative quantum field
theory (QFT) approaches are concerned.  This problem in itself,
however, does not imply that GR cannot be quantized.  For example, the
quantum theories of certain mini-superspace sectors of GR are
known. Unfortunately, since these sectors correspond to quantum
mechanical systems, they do not reveal much about what might be
interesting properties of quantum gravity. Indeed, it may be argued
that quantum mechanical reductions can reveal nothing significant
about the full underlying quantum theory of gravity.

A next attempt at a starting point, short of full GR or
supergravity, might be midi-superspace models -- reductions of GR
which are still field theories.  One such reduction is obtained by
imposing two Killing field symmetries. This reduction has been
extensively studied with a view to quantization
\cite{2kf}.  The most recent work is a claim that
this dimensionally reduced theory can be completely quantized by
finding a representation of the complete classical observable
algebra\cite{ks}. Although apparently complete mathematically, this
quantization has so far given little physical insight into the
underlying quantum theory. It is therefore important to probe such
models a bit further.

In this work we attempt to extract physical consequences from a path
integral quantization of the two-Killing field reduction of GR. The
model is a two dimensional field theory containing two local degrees
of freedom which interact non-linearly. Non-perturbative
considerations reveal that the quantum theory contains a phase
transition. The order parameters for the transition are composite
fields made from spacetime metric functions. We find that the 
composite fields are massless when the ultraviolet (UV) cutoff 
is below a critical energy scale, and are massive when it is 
above this scale. 

Our starting point is the spacetime metric 
\be
ds^2 = e^{2A}\ (-dt^2 + dz^2) + g_{ab}\ dx^adx^b\ , 
\label{t3}
\ee
where $A=A(t,z)$, $x^a = (x^1,x^2)$, and the $2\times 2$ metric 
$$
g_{ab} = R \left( \matrix{ {\rm cosh}W + {\rm cos}\Phi\ {\rm sinh}W &
                             {\rm sin}\Phi\ {\rm sinh}W \cr
               {\rm sin}\Phi\ {\rm sinh}W & 
               {\rm cosh}W - {\rm cos}\Phi\ {\rm sinh}W } \right)
$$
is parameterized by the three functions $R(t,z)$, $\Phi(t,z)$ and
$W(t,z)$. The metric (\ref{t3}) has two commuting space-like Killing
vector fields $\p_{x^1}$ and $\p_{x^2}$, whose orbits have the topology 
of $T^2$.   The $t,z$ coordinates have ranges $0 < t < \infty$ and 
$-\infty\le z \le \infty$. With these conventions, the metric is 
that of the Schmidt spacetime \cite{Sc96}. 
The vacuum
Einstein equations in the gauge $R(t,\theta)=t$, give the two coupled 
two-dimensional evolution equations
\bea \ddot{W} + {1\over t} \dot{W} -
W'' + {\rm sinh}W\ {\rm cosh}W\ ( \Phi'^2 - \dot{\Phi}^2 ) &=& 0\ ,
\label{evo1}\\
\ddot{\Phi} + {1\over t} \dot{\Phi} -\Phi'' 
+ 2\ {{\rm cosh}W\over {\rm sinh}W}\ (\dot{\Phi}\dot{W} -\Phi'W') 
&=& 0\ ,
\label{evo2}
\eea
for  $W(t,\theta)$ and $\Phi(t,\theta)$, and the two ``constraint'' 
equations
\bea
\dot{A} + {1\over 4t} - {t\over 4}\ 
[\ \dot{W}^2 + W'^2  + {\rm sinh}^2W\  (\dot{\Phi}^2 +\Phi'^2)\ ] 
&=& 0\ ,\label{gc1}\\
A' - {t\over 2}\ (\dot{W} W' + {\rm sinh}^2W\ \dot{\Phi}\Phi' ) 
&=& 0\ \label{gc2}. 
\eea
The evolution equations involve $W(t,\theta)$ and
$\Phi(t,\theta)$ and, given a solution to these equations, 
$A(t,\theta)$ is obtained by  integration of the
constraint equations (\ref{gc1}-\ref{gc2}). 

The evolution equations  (\ref{evo1}--\ref{evo2}) may be derived from 
a two-dimensional $\sigma-$model-like action 
\be S_2 (W,\Phi) = {1\over 2} \int dt d\theta\ t\ \sqrt{-\eta}\
\eta^{ab}G_{AB}(Y)\ \p_a Y^A\p_b Y^B\ ,
\label{SG}
\ee
where $Y^1=W$, $Y^2 =\Phi$, $\eta^{ab}={\rm diag}(-,+)$,
$a,b,\cdots=t,z$, and $G_{AB}(Y)\ dY^A dY^B= dW^2 + {\rm
sinh}^2W\ d\Phi^2$ is the unit hyperboloid metric. 
 The $t$
factor in the integrand cannot be absorbed by a 
rescaling of fields, or by introducing a curved two-dimensional 
metric. However this can be done if the model is embedded in 
three dimensions (see below).

Equations (\ref{evo1}-\ref{evo2}) can also be derived 
from the standard three-dimensional $SL(2,\R)$ non-linear \
$\sigma$-model \cite{av},
$$
S_3[X^i,\lambda,\mu^i] =  \int_M \!\!dt dx dz\ 
\sqrt{-\eta} \left\{ \eta^{ab}g_{ij} \p_a X^i\p_b X^j
+ \lambda \left(g_{ij} X^iX^j + 1\right) 
+ \mu^i g_{ij} D X^j\right\}, 
\label{action}
$$
where $g_{ij}={\rm diag}(+,+,-)$, $\eta_{ab}dx^adx^b = -dt^2 + dx^2 +
dz^2$, $\lambda(t,x,z)$ and $\mu^i(t,x,z)$ are Lagrange multiplier
fields, $ D = t \partial_x + x \partial_t$, and $i=1,\cdots, 3$.  The
Lagrange multiplier $\mu^i$ enforces the appropriate reduction to two
dimensions, and the multiplier $\lambda$ enforces the constraints
which give $X^i=X^i(W,\Phi)$. 

For quantization there is the option of using the two
or the three-dimensional action given above. Using the latter allows
inclusion of quantum fluctuations in all three dimensions, whereas the
former restricts these to two dimensions.  Thus we expect on
intuitive grounds that the quantum theories obtained via these two
routes will not be the same.  Because of the richer range for quantum
fluctuations and the standard $\sigma-$model approach it allows, we
use the three-dimensional action $S_3$.

A quantum theory may be defined by the
path integral
\be 
   Z = \int [dX][d\lambda][d\mu]~ {\rm e}^{-iM_P S_3[X^i,\lambda,\mu^i]}\,
\label{pi}
\ee
where $M_P$ is the Planck mass. In this path integral, it is possible to 
perform the Gaussian integral over all the dynamical fields $X^i$. This 
leads to the quantum effective action for the fields $\lambda$ and $\mu^i$. 
 
The saddle point evaluation of the remaining path integral, in its regime 
of validity, gives a first non-perturbative approximation for the quantum 
dynamics of $X^i$. This requires solutions of the 
Euler-Lagrange (EL) equations for $\lambda$ and $\mu^i$ derived from the 
quantum effective action. It is here that we find evidence of a 
phase transition. 

The quantum effective action obtained after integrating over the fields 
$X^i$ is 
\be 
  S_3^{eff} = -{3\over 2}\ {\rm Tr\ ln}\left( \Box +\lambda \right) 
-iM_p\int dtdxdz\ \left[\lambda 
- {1\over 4}\ D\mu^i \left(\Box + \lambda \right)^{-1} D\mu^i \right], 
\ee
where $\Box= -\eta^{ab}\p_a\p_b$ is the three-dimensional flat space 
Laplacian.  The EL equations for $\lambda$ and $\mu^i$
are well defined only in the presence of an UV
cutoff, $\Lambda$, due to a trace term which appears in one of the 
equations.   
At this stage therefore, a second energy scale, $\Lambda$, enters 
our quantum model; all solutions of the EL equations  depend on 
the cutoff. The simplest solutions are
\be 
\lambda = \lambda^* ={\rm const.}\ \ \ \   {\rm and}\ \ \ \  
\mu^i = \mu^{i*}={\rm const.}.
\ee
These particular solutions exist {\it only if} $\Lambda \ge \alpha M_P$ 
\cite{poly}, where $\alpha$ is dimensionless constant of order unity.
 
What do these solutions imply for the phase of the theory? In the 
saddle point approximation, the two-point function is  
$$
\langle X^i(x) X^j(y) \rangle \sim g^{ij}
\int^\Lambda {d^3k\over (2\pi)^3} {e^{ik(x-y)}\over k^2 - \lambda^*},
$$ 
This simple result demonstrates that the $X^i$ fields are massive
with mass $\lambda^*$ on scales $\Lambda > \alpha M_P$, and so 
we can speak of a ``massive phase'' in this regime. A careful treatment 
reveals that 
\be 
\lambda^*(\Lambda, M_P) = 
{\rm const.}~ {\Lambda^4\over \alpha^2} 
\left({1\over M_P} - {\alpha\over \Lambda}\right)^2,
\ee
so that the mass is zero at $\Lambda = \alpha M_P$. 
  
It is surprising that integrating out all the $X^i$ fields in
eqn. (\ref{pi}) leads to these solutions of the effective action only
if $\Lambda\ge \alpha M_P$. Is it possible to extend this solution
to the region $\Lambda< \alpha M_P$? Indeed, it is.  One must go
back to the original path integral and assume that one of the $X^i$
fields is a constant, that is, it has a non-zero vacuum expectation
value (vev). This modifies the EL equations for $\lambda$
and $\mu^i$, such that now $\lambda=0$ and $\mu^i =
\mu^{i*}={\rm constant}$ are indeed solutions, (while the vev is found to 
depend on $\Lambda$ and $M_P$). Thus, in the $\Lambda < \alpha M_P$
parameter regime, we are in a ``massless phase,'' and there is
spontaneous breaking of the $SL(2,\R)$ symmetry (due to the direction
picked out by the constant component of the $X^i$ field). These
results are summarized in Figure 1.

\begin{figure}[t]
\epsfxsize=120mm
\hspace*{20mm}\epsffile {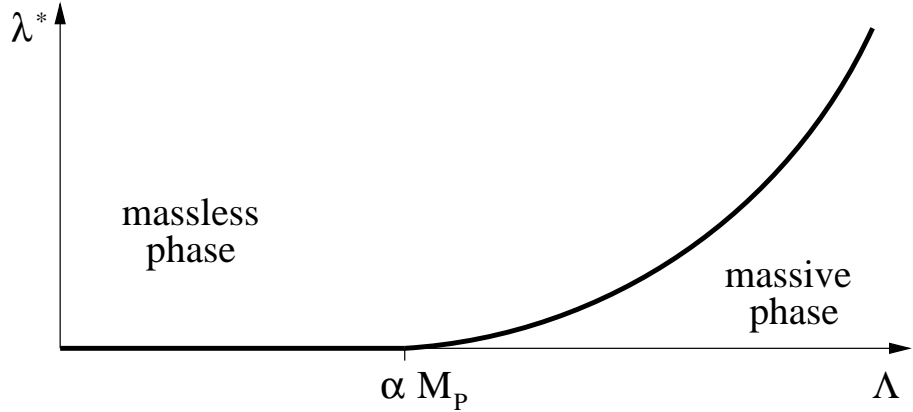}
\vspace*{1mm}
\caption{Spontaneous symmetry breaking occurs for $\Lambda<\alpha M_P$ 
and gravitational excitations become massless.} 
\end{figure}

The steps we have outlined are standard for exploring the possibility of 
phase transitions in QFT. What is surprising, as we have 
seen, is that an analogous treatment applies to the path integral (\ref{pi}) 
for reduced Einstein gravity. {\it This appears to be the first concrete 
indication from GR itself that a phase transition  
occurs at very short distances.} 

What do these results imply for quantum gravity? Perhaps the main
insight is that gravitational excitations are massive 
above the Planck energy scale, and massless, with spontaneous symmetry 
breaking, below this scale. The symmetry breaking below $M_P$ in this 
reduction of GR is relatively slight --- it is a breaking of global 
$SL(2,\R)$. Nevertheless,  it is significant in that it occurs at all. 

It is important to emphasize that this phenomena is essentially due 
to the two non-linearly interacting degrees of freedom in this model; 
it is missed if only one field is present. The implication of our 
result for full quantum gravity is the  possibility that 
the graviton is a massless Goldstone boson associated with spontaneous 
symmetry breaking at the Planck scale. 

Several questions remain for further research: Is the dimensionally
reduced model considered here renormalizable? If so what is the 
$\beta$--function and its fixed points (if any)? What is the spacetime 
metric at scales below the Planck length?

With respect to the first question, it is well known that the regular
$\sigma-$model in three dimensions is renormalizable. Therefore it is
clear that this must be the case also for the present model, even with
the dimensional reduction constraint it contains. Surprisingly,
however, this result is not straightforward to establish if the
constraint is first solved classically and an action in two dimensions
is used as the starting point. This is because the resulting
$\sigma-$model like action has an explicit time factor in the
integrand, which results in the kinetic operator not being
Sturm--Lioville. This crucially affects the standard evaluation of the
Gaussian integral. Thus, establishing renormalizability and
calculating the $\beta$--function is straightforward from the
three-dimensional perspective, but not from the two-dimensional
one. Indeed the model may not even be renormalizable from the latter
perspective, where the $\sigma$ model constraint is explicitly time
dependent.  If so, this would be yet another indication of the
important differences between dimensionally reducing classically and
then quantizing, versus quantizing by including the constraint quantum
mechanically as we have done here.

\end{document}